\documentclass[11pt, draftcls, onecolumn]{IEEEtran}
\usepackage{cite,bbm,graphicx,amsmath,amssymb,mathrsfs,epsf} \usepackage{multirow}
\usepackage{subfigure,cite,graphicx,epsfig,amsmath,amssymb,mathrsfs,epsf,graphics,color,enumerate}

\newtheorem{theorem}{Theorem}

\newtheorem{remark}{Remark}

\def \E{\operatorname{E}}

\begin{document}
\title{The Degraded Gaussian Diamond-Wiretap Channel}
\author{\authorblockN{Si-Hyeon Lee and Ashish Khisti\\}
\authorblockA{Electrical and Computer Engineering, University of Toronto, Toronto, Canada\\
Email: sihyeon.lee@utoronto.ca, akhisti@comm.utoronto.ca }\thanks{This work was supported by QNRF, a member
 of Qatar Foundation, under NPRP project 5-401-2-161.}} \maketitle

\maketitle

\begin{abstract}
In this paper, we present nontrivial upper and lower bounds on the secrecy capacity of  the degraded Gaussian diamond-wiretap channel and identify several ranges of channel parameters where these bounds coincide with useful intuitions. Furthermore, we investigate the effect of the presence of an eavesdropper on the capacity. We consider the following two scenarios regarding the availability of randomness:  1) a common randomness is available at the source and the two relays and 2) a randomness is available only at the source and there is no available randomness at the relays.  We obtain the upper bound by taking into account the correlation between the two relay signals and the availability of randomness at each encoder. For the lower bound, we propose two types of coding schemes: 1) a decode-and-forward scheme where the relays cooperatively transmit the message and the fictitious message and 2) a partial DF scheme incorporated with multicoding in which each relay sends an independent partial message and the whole or partial fictitious message using dependent codewords. 
\end{abstract}

\IEEEpeerreviewmaketitle
\section{Introduction}
The diamond channel introduced by Schein \cite{schein_thesis} consists of a broadcast channel (BC) from a source to two relays and a multiple access channel (MAC) from the two relays to a destination. The capacity of the diamond channel is not known in general. To simplify the problem, let us consider a diamond channel having BC with two orthogonal links and  Gaussian MAC. In this setup, there is a tension between increasing the amount of information sent over the BC and increasing the coherent combining gain for the MAC. Two coding schemes corresponding to the extremes would be partial decode-and-forward, where independent partial messages are sent to the relays, and decode-and-forward (DF), where the whole message is sent to each of the relays. By incorporating multicoding at the source, \cite{TraskovKramer:07}, \cite{KangLiu:11} proposed a coding scheme in which the relays send independent partial messages using dependent codewords and showed that this coding scheme strictly outperforms the DF and partial DF in some regime. Furthermore, \cite{KangLiu:11} showed an upper bound by taking into account the correlation between the two relay signals, which is strictly tighter than the cutset bound.  This upper bound was shown to coincide with the lower bound of \cite{TraskovKramer:07}, \cite{KangLiu:11} for some channel parameters. 

\begin{figure}[t]
 \centering
  {
  \includegraphics[width=85mm]{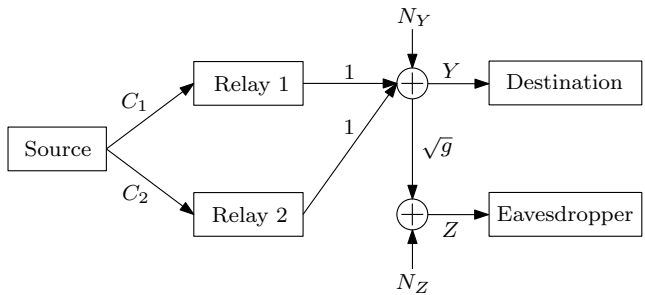}}
  \caption{Physically degraded diamond-wiretap channel} \label{fig:physical}
\end{figure}

In this paper, we consider the degraded Gaussian diamond-wiretap channel presented in Fig. \ref{fig:physical} and present lower and upper bounds on the secrecy capacity by exploiting the correlation between the two relay signals. We identify several ranges of channel parameters where these bounds coincide with useful intuitions and investigate the effect of the presence of an eavesdropper on the capacity.  We note that this model is a natural first step to studying  diamond-wiretap channel because the sum secrecy capacity of the multiple access-wiretap channel has been characterized only for the degraded Gaussian case \cite{TekinYener:08}. A practical situation corresponding to this model is the side channel attack \cite{Kocher:99} where the eavesdropper attacks by probing the physical signals such as timing information and  power consumption  leaked from the legitimate destination.  In the presence of an eavesdropper, the technique of utilizing randomness is widely used to confuse the eavesdropper. We consider the following two scenarios regarding the availability of randomness:  1) a common randomness of rate $R'$ is available at the source and the two relays and 2) a randomness of rate $R'$ is available only at the source and there is no available randomness at the relays. See  \cite{KobayashiYamamotoOgawa:11}, \cite{ChouBlock:14} for the related works assuming restricted randomness at encoders.

For the upper bound, we generalize the upper bound on the capacity of the diamond channel \cite{KangLiu:11} and the upper bound on the sum secrecy capacity of the multiple access-wiretap channel \cite{TekinYener:08}.       
For the lower bound, we propose two types of coding schemes: 1) a decode-and-forward (DF) scheme where the relays cooperatively transmit the message and the fictitious message and 2) a partial DF scheme incorporated with multicoding in which each relay sends an independent partial message and the whole or partial fictitious message using dependent codewords. If there is no secrecy constraint, our partial DF scheme incorporated with multicoding falls back to that in \cite{TraskovKramer:07}, \cite{KangLiu:11}. Interestingly, in the presence of the eavesdropper, the availability of randomness at the encoders is shown to affect the optimal selection of correlation coefficient between the two relay signals in our proposed schemes. 

The remaining  this paper is organized as follows. In Section \ref{sec:model}, we formally present the model of the degraded Gaussian diamond-wiretap channel. Our main results on the secrecy capacity are given in Section \ref{sec:main}. In Section \ref{sec:proof}, we derive our upper and lower bounds on the secrecy capacity. We conclude this paper in Section \ref{sec:conclusion}.

\section{Model} \label{sec:model}
Consider the degraded Gaussian diamond-wiretap channel in Fig.  \ref{fig:physical} that consists of a source, two relays, a legitimate destination, and an eavesdropper. The source is connected to two relays through orthogonal links of capacities $C_1$ and $C_2$ and there is no direct link from the source to the legitimate destination or eavesdropper. The channel outputs $Y$ and $Z$ at the legitimate destination and the eavesdropper, respectively, are given as $Y=X_1+X_2+N_Y$ and $Z=\sqrt{g}Y+N_Z$, where $g\in [0,1)$, $X_1$ and $X_2$ are the channel inputs from relay 1 and relay 2, respectively, $N_Y$ is the Gaussian noise with zero mean and unit variance at the legitimate destination, and $N_Z$ is the Gaussian noise with zero mean and variance of $1-g$ at the eavesdropper. $N_Y$ and $N_Z$ are assumed to be independent. The transmit power constraint at relay $k=1,2$ is given as $\frac{1}{n}\sum_{i=1}^nX_{k,i}^2\leq P_k$, where $n$ denotes the number of channel uses. Note that the channel output at the eavesdropper is a physically degraded version of the channel output at the legitimate destination.

We consider the following two scenarios regarding the availability of randomness. In the first scenario, a common fictitious message $M$ of rate $R'$, i.e., $M\sim \mbox{Unif}[1:2^{nR'}]$\footnote{$[i:j]$ for two integers $i$ and $j$ denotes the set $\{i,i+1,\ldots,j\}$.} is available at the source and the two relays. In this case, a $(2^{nR}, n)$ secrecy code consists of a message $W\sim \mbox{Unif}[1:2^{nR}]$, an encoding function at the source that maps $(W,M)\in [1:2^{nR}]\times [1:2^{nR'}]$ to $(J_1, J_2)\in [1:2^{nC_1}]\times [1:2^{nC_2}]$, an encoding function at relay $k=1,2$ that maps $(J_k,M)\in [1:2^{nC_k}]\times [1:2^{nR'}]$ to $X_k^n \in \mathcal{X}_k^n$, and a decoding function at the legitimate destination that maps $Y^n\in \mathcal{Y}^n$ to $\hat{W} \in [1:2^{nR}]$. In the second scenario, a fictitious message $M$ of rate $R'$  is available only at the source and the encoding at the two relays is restricted to be deterministic. In this case, the encoding function at relay $k=1,2$ maps $J_k\in [1:2^{nC_k}]$ to $X_k^n \in \mathcal{X}_k^n$.

For both scenarios, the probability of error is given as  $P_e^{(n)}=P(\hat{W}\neq W)$. A secrecy rate of $R$ is said to be achievable if there exists a sequence of $(2^{nR},n)$ codes such that 
$\lim_{n\rightarrow \infty}P_e^{(n)}=0$ and $\lim_{n\rightarrow \infty}\frac{1}{n}I(W;Z^n)=0$. The secrecy capacity is the supremum of all achievable secrecy rates. Let $C_S^{(1)}$ and $C_S^{(2)}$ denote the secrecy capacity for the first scenario and for the second scenario, respectively. 

\begin{figure}[t]
 \centering
  {
  \includegraphics[width=85mm]{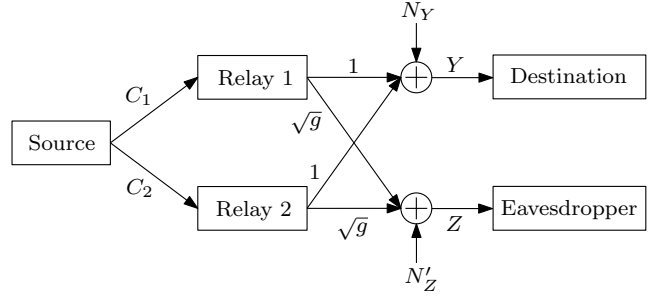}}
  \caption{Stochastically degraded diamond-wiretap channel} \label{fig:stochastic}
\end{figure}

\begin{remark}
Because the legitimate destination and the eavesdropper do not cooperate, the secrecy capacity in Fig.  \ref{fig:physical} is the same as that of stochastically degraded case in Fig. \ref{fig:stochastic}, in which $Z$ is given as $Z=\sqrt{g}X_1+\sqrt{g}X_2+N_Z'$, where $N_Z'$ has zero mean and unit variance and is independent of $N_Y$. 
\end{remark}
 
\section{Main Results} \label{sec:main} 
In this section, we present main results of this paper on the secrecy capacity of the degraded Gaussian diamond-wiretap channel described in Section \ref{sec:model}. For the brevity of presentation, let us define the following functions:
\begin{subequations}\allowdisplaybreaks
\begin{align}
f_1(\rho)&=C_1+\frac{1}{2}\log(1+(1-\rho^2)P_2)\\
f_2(\rho)&=C_2+\frac{1}{2}\log(1+(1-\rho^2)P_1)\\
f_3(\rho)&=C_1+C_2-\frac{1}{2}\log(\frac{1}{1-\rho^2}) \\
f_4(\rho)&=\frac{1}{2}\log (1+P_1+P_2+2\rho\sqrt{P_1P_2})\\
f_5(\rho)&=\frac{1}{2}\log (1+g(P_1+P_2+2\rho\sqrt{P_1P_2}))\\
f_6(\rho)&=\frac{1}{2}\log \left(\frac{1+g(P_1+P_2+2\rho\sqrt{P_1P_2})}{1+g(1-\rho^2)P_2}\right)\\
f_7(\rho)&=\frac{1}{2}\log \left(\frac{1+g(P_1+P_2+2\rho\sqrt{P_1P_2})}{1+g(1-\rho^2)P_1}\right),
\end{align} \label{eqn:f_def}
\end{subequations}
where the domain of $f_1,f_2, f_3$, $f_6$, and $f_7$ is $[-1,1]$ and that of $f_4$ and $f_5$ is $[-\bar{\rho}, 1]$ for $\bar{\rho}=\frac{P_1+P_2}{2\sqrt{P_1P_2}}$.\footnote{By convention, we assume that $f_3(\rho)$ becomes negative infinity when $|\rho|=1$.}

The following two theorems give upper and lower bounds on $C_S^{(1)}$, respectively, whose proofs are in Section \ref{sec:proof}.
\begin{theorem} \label{thm:common_ub} For $R'\geq 0$, $C_S^{(1)}$ is upper-bounded by  
\[\min(\max(S_1, S_2), \max(S_3, S_4)),\] where 
\begin{align*}
S_1&=\max_{0\leq \rho \leq \rho^*} \min(f_1(\rho), f_2(\rho), f_3(\rho), f_4(\rho)) \\
S_2&=\max_{\rho^*< \rho \leq 1} \min(f_1(\rho), f_2(\rho), f_3(0), f_4(\rho))\\
S_3&=\max_{0\leq \rho \leq \rho^*} \min(f_1(\rho), f_2(\rho), f_3(0), \frac{f_3(\rho)+f_4(\rho)}{2},f_4(\rho)-f_5(\rho))\\
S_4&=\max_{\rho^*<\rho \leq 1} \min(f_1(\rho), f_2(\rho), f_3(0), f_4(\rho)-f_5(\rho)) 
\end{align*} 
for  $\rho^*=\sqrt{1+\frac{1}{4P_1P_2}}-\frac{1}{2\sqrt{P_1P_2}}$. We note that the functions $f_k$'s for $k\in [1:5]$ are defined in (\ref{eqn:f_def}).
\end{theorem}
 
\begin{theorem} \label{thm:common_lb} For $\rho\in [-1,1]$ and $R'\geq f_5(\rho)$, $C_S^{(1)}$ is lower-bounded by  
\begin{align*}
\max(R_{\mathrm{DF}}^{(1)}(\rho), R_{\mathrm{PDF-M}}^{(1)}(\rho)), 
\end{align*}
where 
\begin{align*}
R_{\mathrm{DF}}^{(1)}(\rho)&=\min (C_1, C_2, f_4(\rho)-f_5(\rho)) \\
R_{\mathrm{PDF-M}}^{(1)}(\rho)&=\min(f_1(\rho), f_2(\rho), f_3(\rho), f_4(\rho)-f_5(\rho)).
\end{align*}
We note that the functions $f_k$'s for $k\in [1:5]$ are defined in (\ref{eqn:f_def}).
\end{theorem}

In Theorem \ref{thm:common_ub}, we note that the upper bound $\max(S_1,S_2)$ is the same as that in \cite{KangLiu:11} that assumes no secrecy constraint. This is natural because the secrecy capacity is upper-bounded by the capacity without secrecy constraint, which is not affected by the common randomness at the encoders. To derive the upper bound $\max(S_3,S_4)$, we generalize the bounding techniques \cite{KangLiu:11} and \cite{TekinYener:08} taking into account the secrecy constraint and  the available randomness at the encoders.  

In Theorem \ref{thm:common_lb}, $R_{\mathrm{DF}}^{(1)}(\rho)$ is achieved by using a DF scheme where the source sends the message to both relays and the relays cooperatively transmit the message and the common fictitious message over the wiretap channel. On the other hand, $R_{\mathrm{PDF-M}}^{(1)}(\rho)$ is achieved by a partial DF incorporated with multicoding (PDF-M) where each relay sends an independent partial message and the common fictitious message using dependent codewords. The source performs multicoding as follows: the message $w$ is represented as two partial messages $(w_1, w_2)$, a codebook for relay $k=1,2$ consisting of independently generated $x_k^n$ sequences is constructed for each $w_k$ and $m$, and the source finds a jointly typical sequence pair $(x_1^n(w_1,m,l_1), x_2^n(w_2,m,l_2))$ and sends $(w_k,l_k)$ to relay $k$ for $k=1,2$. A more detailed explanation for the PDF-M scheme is given in Section \ref{sec:proof}. Let $R_{\mathrm{PDF}}^{(1)}=R_{\mathrm{PDF-M}}^{(1)}(0)$ denote the partial DF (PDF) rate without multicoding at the source.

\begin{figure}[t]
 \centering\subfigure[]
  {\includegraphics[width=120mm]{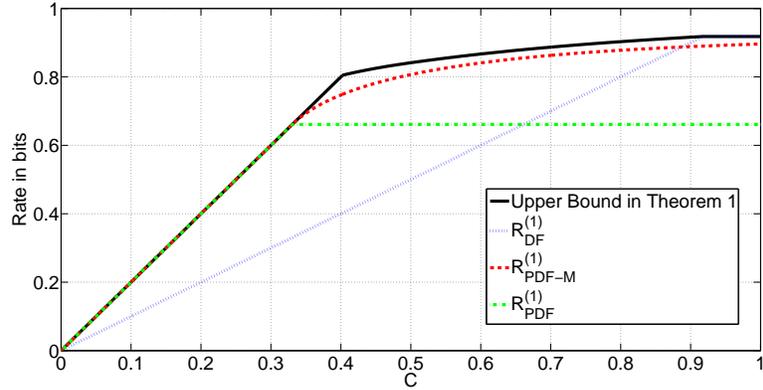}}
  \hspace{0.2in}
 \subfigure[]
  {\includegraphics[width=120mm]{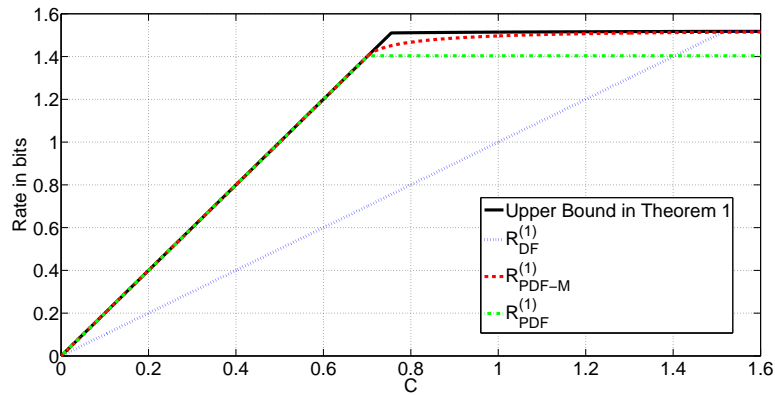}}\\
 \caption{Bounds for the first scenario when (a) $P=1, g=0.1$ and (b) $P=10, g=0.1$. } \label{fig:common}
\end{figure}

To compare our lower and upper bounds, let us consider sufficiently large $R'$ and symmetric channel parameters, i.e., $P_1=P_2=P$ and $C_1=C_2=C$ for some nonnegative $P$ and $C$. It can be easily proved that 1) the PDF scheme, which achieves\footnote{For $P_1=P_2$, $C_1=C_2=C$, and $\rho=0$, $f_1(0)$ and $f_2(0)$ become redundant.} $\min(f_3(0), f_4(0)-f_5(0))$, is optimal for $C\leq\frac{1}{2}(f_4(0)-f_5(0))$, i.e., the BC cut is the bottleneck, and 2) the DF scheme, which achieves $\min (C, f_4(1)-f_5(1))$, is optimal for $C\geq f_4(1)-f_5(1)$,  i.e., the MAC cut is the bottleneck. When neither the BC cut nor the MAC cut is the bottleneck, the PDF-M scheme strictly outperforms the PDF and DF schemes for some range of $C$ as shown in Fig. \ref{fig:common}. For example, when $P=1$ and $g=0.1$, the PDF-M scheme strictly outperforms the PDF and DF schemes for $0.33<C<0.89$. Furthermore, Fig.  \ref{fig:common} shows that the PDF bound gets close to the upper bound in Theorem \ref{thm:common_ub} as $P$ increases. The following theorem states that the PDF scheme is indeed asymtotically optimal as $P_1$ or $P_2$ tends to infinity, whose proof is relegated to the end of this section.

\begin{theorem} \label{thm:asymptotic}
For the first scenario with $R'\geq f_5(0)$ and $P_1\rightarrow \infty$ or $P_2\rightarrow \infty$,\footnote{$C_1$ and $C_2$ are not necessarily fixed and can be arbitrary functions of $P_1$ and $P_2$.} the PDF scheme is asymptotically optimal. 
\end{theorem}

Next, the following two theorems give upper and lower bounds on $C_S^{(2)}$, respectively, whose proofs are in Section \ref{sec:proof}.

\begin{theorem} \label{thm:no_ub}
For $R'\geq 0$, $C_S^{(2)}$ is upper-bounded by   \[\max(T_1, T_2, T_3),\] where 
\begin{align*}
T_1&=\max_{-\bar{\rho}\leq \rho < 0} \min(f_1(0), f_2(0), f_3(0), f_4(\rho))-f_5(\rho) \\
T_2&=\max_{0\leq \rho \leq \rho^*} \min(f_1(\rho), f_2(\rho), f_3(\rho), f_4(\rho))-f_5(\rho)\\
T_3&=\max_{\rho^*< \rho \leq 1} \min(f_1(\rho), f_2(\rho), f_3(0), f_4(\rho))-f_5(\rho).
\end{align*} 
We note that the functions $f_k$'s for $k\in [1:5]$ are defined in (\ref{eqn:f_def}), $\bar{\rho}=\frac{P_1+P_2}{2\sqrt{P_1P_2}}$, and $\rho^*=\sqrt{1+\frac{1}{4P_1P_2}}-\frac{1}{2\sqrt{P_1P_2}}$.
\end{theorem}
 
\begin{theorem} \label{thm:no_lb}
For $\rho\in [-1,1]$ such that $R'\geq f_5(\rho)$, $C_S^{(2)}$ is lower-bounded by 
\begin{align*}
\max(R_{\mathrm{DF}}^{(2)}(\rho), R_{\mathrm{PDF-DF-M}}^{(2)}(\rho), R_{\mathrm{PDF-PDF-M}}^{(2)}(\rho)),
\end{align*}
where 
\begin{align*}
R_{\mathrm{DF}}^{(2)}(\rho)&=\min(C_1, C_2, f_4(\rho))-f_5(\rho)\\
R_{\mathrm{PDF-DF-M}}^{(2)}(\rho)&=\min(f_1(\rho), f_2(\rho), f_3(\rho)-f_5(\rho), f_4(\rho))-f_5(\rho)\\
R_{\mathrm{PDF-PDF-M}}^{(2)}(\rho)&=(\min(f_1(\rho), f_2(\rho), f_3(\rho), f_4(\rho))-f_5(\rho))\cdot \mathbbm{1}_{C_1>f_6(\rho), C_2>f_7(\rho)}.
\end{align*}
We note that the functions $f_k$'s for $k\in [1:7]$ are defined in (\ref{eqn:f_def}).
\end{theorem}
Note that in both the upper and lower bounds for the first scenario, the term $f_5(\rho)$, which corresponds to the required rate of randomness to confuse the eavesdropper, appears only with $f_4(\rho)$, which signifies the amount of information sent through the MAC. In contrast, in both the upper and lower bounds for the second scenario, because the fictitious message has to be sent through the BC, $f_5(\rho)$ appears in common for all terms. This affects sufficient ranges of correlation coefficient for the lower bounds for large enough $R'$  as remarked in the following. 

\begin{remark} \label{remark:negative}
For large enough $R'$, sufficient ranges of correlation coefficient $\rho$ for the lower bounds in Theorem \ref{thm:common_lb} and Theorem \ref{thm:no_lb} are different. 
For the first scenario, note that the DF rate is maximized at $\rho=1$ and that it is enough to consider nonnegative $\rho$ for the PDF-M scheme. On the other hand, for the second scenario, because the minus term $-f_5(\rho)$ is common for all terms, considering smaller $\rho$ can be beneficial by decreasing $f_5(\rho)$ and we need consider all $-1\leq \rho \leq 1$. 
\end{remark}

\begin{figure}[t]
 \centering
  {\includegraphics[width=120mm]{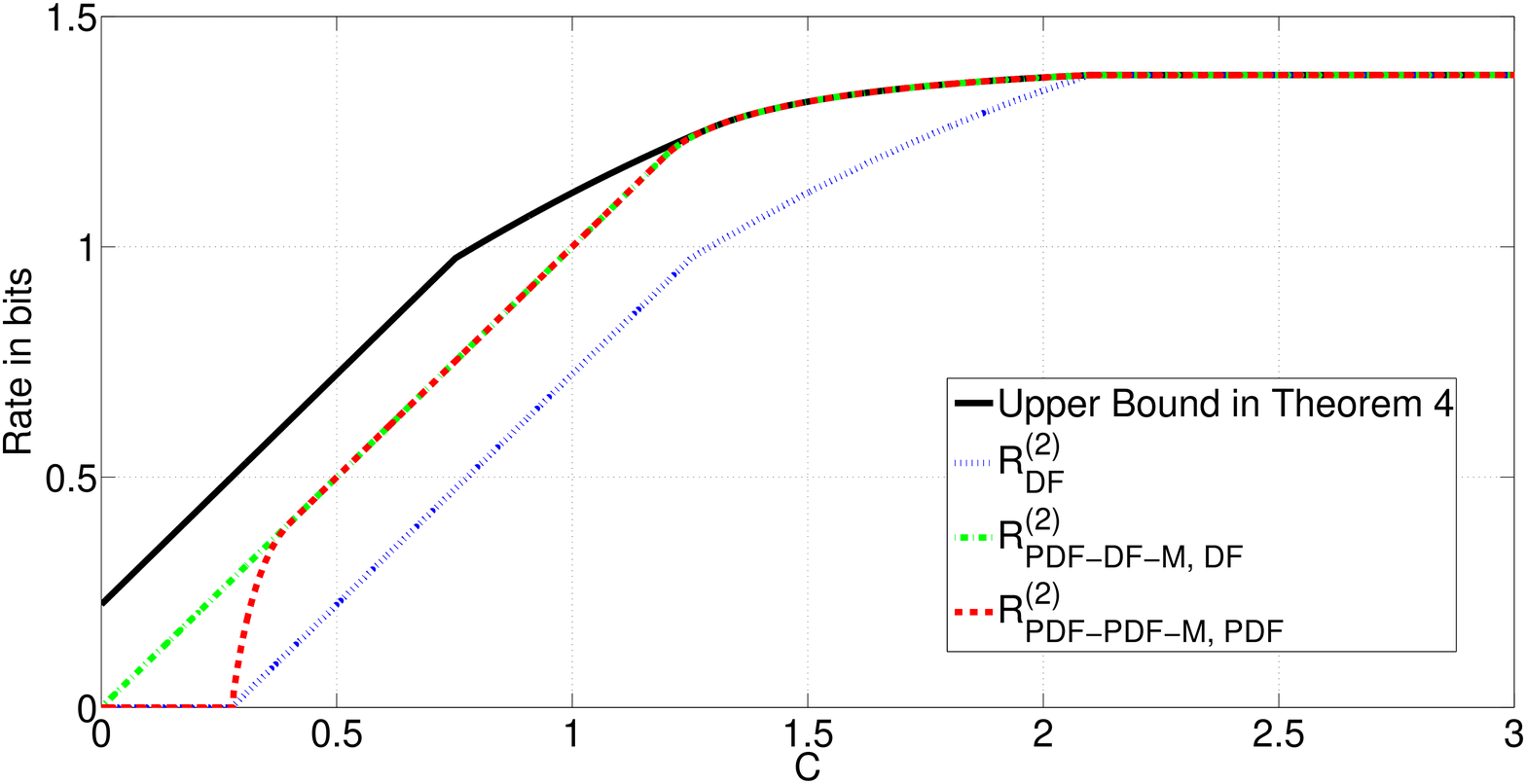}}
 \caption{For sufficiently large $R'$, $g=0.1$, $C_1=C$, $C_2=C+2$, $P_1=10$, and $P_2=1$,  $R_{\mathrm{PDF-DF-M}}^{(2)}(\rho)$ is strictly higher than $R_{\mathrm{PDF-PDF-M}}^{(2)}(\rho)$ for some range of $C$. } \label{fig:PDFDF}
\end{figure}

In the DF scheme for the second scenario, the source sends to both relays the fictitious message as well as the message. Hence, $R_{\mathrm{DF}}^{(2)}$ is obtained from $R_{\mathrm{DF}}^{(1)}$ by replacing $C_1$ and $C_2$ by $C_1-f_5(\rho)$ and $C_2-f_5(\rho)$, respectively. For a partial DF scheme incorporated with multicoding for the second scenario, a straightforward extension from that for the first scenario is to let the source send the fictitious message $m$ as well as the partial message $w_k$ and the relay codeword index $l_k$ to relay $k$ for $k=1,2$. Since each relay decodes a partial genuine message and a whole fictitious message, we call this scheme as PDF-DF-M scheme. Note that $R_{\mathrm{PDF-DF-M}}^{(2)}(\rho)$ is obtained by replacing $C_1$ and $C_2$ by $C_1-f_5(\rho)$ and $C_2-f_5(\rho)$, respectively, in $R_{\mathrm{PDF-M}}^{(1)}$.  However, since the same fictitious message is sent to both relays, there exists inefficiency in the use of the BC. To resolve this inefficiency, we let each of relay codebooks be indexed by independent partial fictitious message, i.e., codebook for relay $k=1,2$ is constructed for each $(w_k, m_k)$ by representing $m$ as two partial fictitious messages $(m_1,m_2)$. By using this PDF-PDF-M scheme where each relay decodes a partial genuine message and a partial fictitious message, we show that $R_{\mathrm{PDF-PDF-M}}^{(2)}(\rho)$ is achievable, which has $f_3(\rho)$ intead of $f_3(\rho)-f_5(\rho)$ in  $R_{\mathrm{PDF-DF-M}}^{(2)}(\rho)$. We note that having independent fictitious message at each relay reduces the achievable rate region over the MAC, which results in additional contraints $C_1>f_6(\rho)$ and $C_2>f_7(\rho)$ in $R_{\mathrm{PDF-PDF-M}}^{(2)}(\rho)$. Nevertheless, as long as $C_1=C_2$,  $R_{\mathrm{PDF-PDF-M}}^{(2)}(\rho)$ is always higher than or equal to $R_{\mathrm{PDF-DF-M}}^{(2)}(\rho)$ because $f_3(\rho)>2f_5(\rho)$, which should be satisfied if $R_{\mathrm{PDF-DF-M}}^{(2)}(\rho)>0$, implies $C_1>f_6(\rho)$ and $C_2>f_7(\rho)$. If $C_1\neq C_2$, $R_{\mathrm{PDF-DF-M}}^{(2)}(\rho)$ can be strictly higher than $R_{\mathrm{PDF-PDF-M}}^{(2)}(\rho)$ as illustrated in Fig. \ref{fig:PDFDF}. Let $R_{\mathrm{PDF-DF}}^{(2)}=R_{\mathrm{PDF-DF-M}}^{(2)}(0)$ and $R_{\mathrm{PDF-PDF}}^{(2)}=R_{\mathrm{PDF-PDF-M}}^{(2)}(0)$ denote the rates of PDF-DF and PDF-PDF schemes (without multicoding).

\begin{figure}[t]
 \centering\subfigure[]
  {\includegraphics[width=120mm]{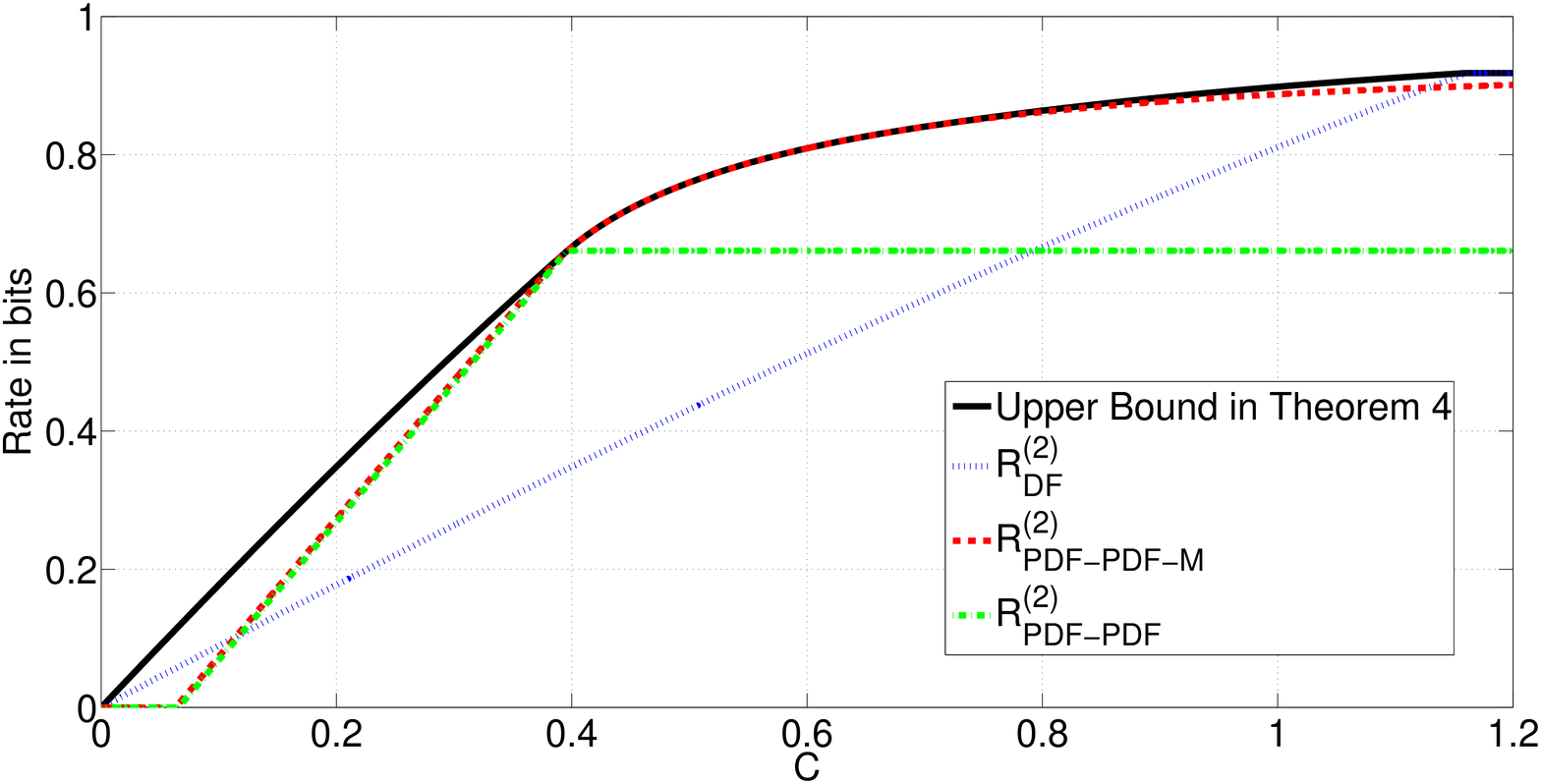}}
  \hspace{0.2in}
 \subfigure[]
  {\includegraphics[width=120mm]{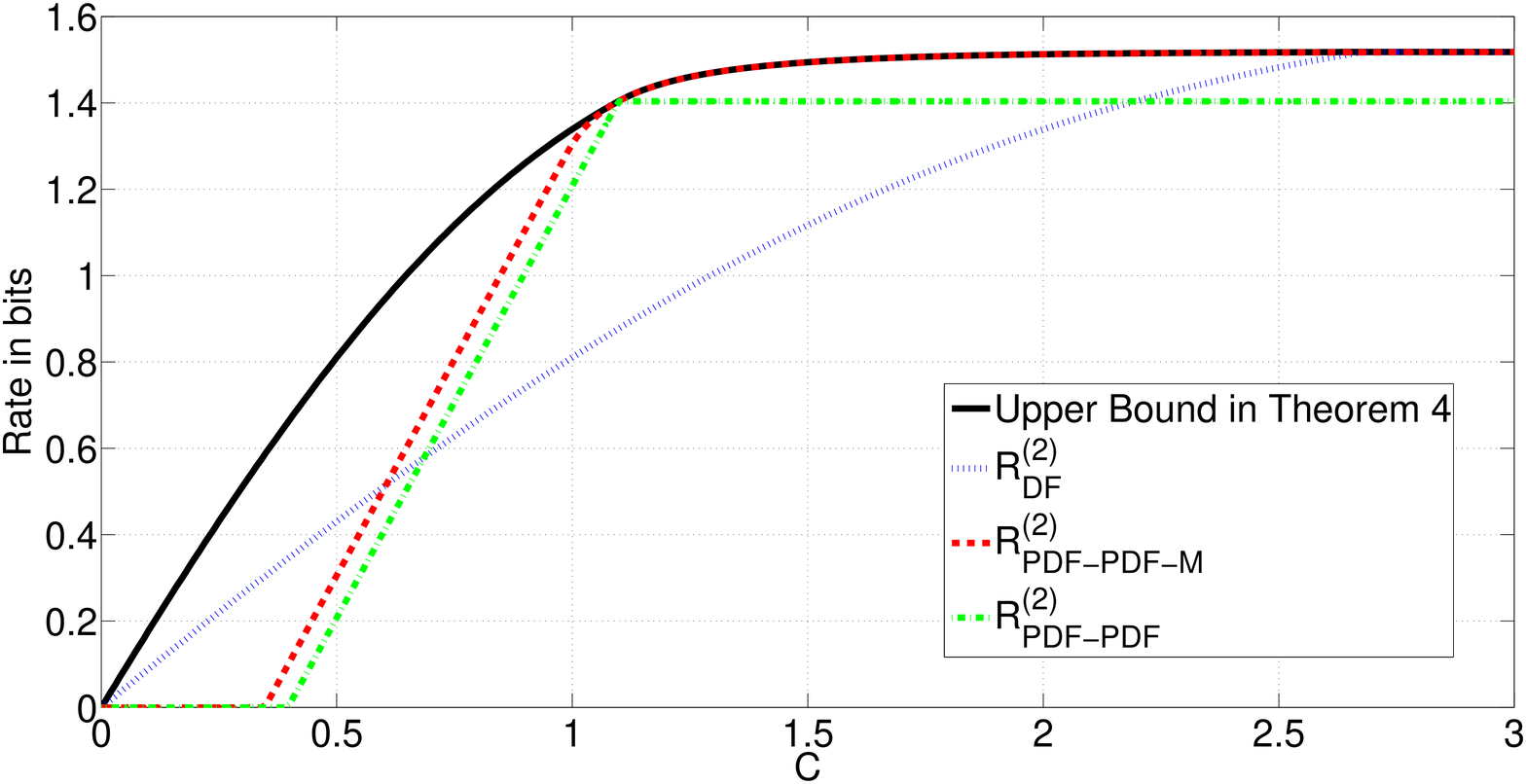}}\\
 \caption{Bounds for the second scenario when (a) $P=1, g=0.1$ and (b) $P=10, g=0.1$. } \label{fig:no}
\end{figure}

Similarly as for the first scenario, let us consider sufficiently large $R'$ and symmetric channel parameters. Since $C_1=C_2$, we only consider the DF, PDF-PDF-M, and PDF-PDF schemes for the lower bounds. It can be easily proved that the DF scheme, which achieves $\max_{\rho\in [-1,1]}\min (C, f_4(\rho))-f_5(\rho)$, is optimal for $C\geq f_4(1)$,  i.e., the MAC cut is the bottleneck. We can see in Fig. \ref{fig:no} that the PDF-PDF  rate coincides with the PDF-PDF-M rate at one point. This is because a negative correlation between the two relay signals is helpful for small $C$ due to the reason in Remark \ref{remark:negative}, i.e., the BC cut is the bottleneck, and positive correlation becomes beneficial as $C$ increases, i.e., the MAC cut becomes bottleneck. Fig. \ref{fig:no} also shows that the PDF-PDF-M rate is zero up to some threshold value of $C$ due to the constraint $C>f_6(\rho)$ in $R_{\mathrm{PDF-PDF-M}}^{(2)}(\rho)$ and the threshold value decreases as $P$ decreases. Indeed, we can prove that the threshold value tends to zero as $P$ tends to zero. Furthermore, Fig. \ref{fig:no} shows that the PDF-PDF-M rate coincides with the upper bound in Theorem \ref{thm:no_ub} for some range of $C$, e.g., $1.1<C<2.18$ when $P=10$ and $g=0.1$. The following theorem  gives a condition where the PDF-PDF-M rate coincides with the upper bound  in Theorems \ref{thm:no_ub}, whose proof is relegated to the end of this section.

\begin{theorem} \label{thm:cap_condition}
For the second scenario with sufficiently large $R'$ and symmetric channel parameters, the PDF-PDF-M rate in Theorem \ref{thm:no_lb} coincide with the upper bound in Theorem \ref{thm:no_ub}, and the secrecy capacity is given as $f_3(\rho')-f_5(\rho')$ for 
\begin{align}
\frac{1}{4}\log (1+2P)\leq C \leq \frac{1}{4}\log (1+2(1+\rho^*)P)+\frac{1}{4}\log (\frac{1}{1-\rho^{*2}}) \label{eqn:cap_condition}
\end{align}
such that that at least one of $f_1(\rho^*)-f_5(\rho^*)\leq f_3(\rho')-f_5(\rho')$ and $f_3(0)-f_5(\rho^*)\leq f_3(\rho')-f_5(\rho')$ is satisfied, where $\rho^*=\sqrt{1+\frac{1}{4P_1P_2}}-\frac{1}{2\sqrt{P_1P_2}}$ and $\rho'\in [0, \rho^*]$ is such that $f_3(\rho')=f_4(\rho')$.\footnote{We note that under the condition (\ref{eqn:cap_condition}), $\rho'\in [0, \rho^*]$ such that $f_3(\rho')=f_4(\rho')$ exists.}
\end{theorem} 

Theorem \ref{thm:cap_condition} indicates that the upper and lower bounds in Theorems \ref{thm:no_ub} and \ref{thm:no_lb} coincide for $1.1<C<2.18$ when $P=10$ and $g=0.1$ and for $1.91<C<3.82$ when $P=100$ and $g=0.1$.

\begin{remark}
For $g=0$, the bounds in Theorems \ref{thm:common_ub}-\ref{thm:no_lb} fall back to those in \cite{KangLiu:11}. 
\end{remark}

\begin{figure}[t]
 \centering\subfigure[]
  {\includegraphics[width=120mm]{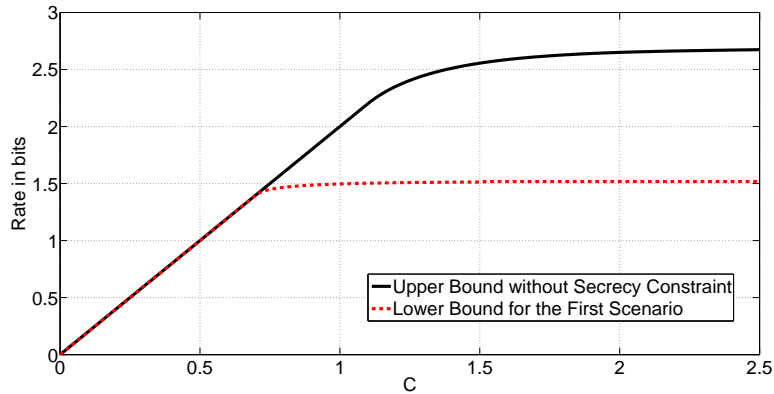}}
  \hspace{0.2in}
 \subfigure[]
  {\includegraphics[width=120mm]{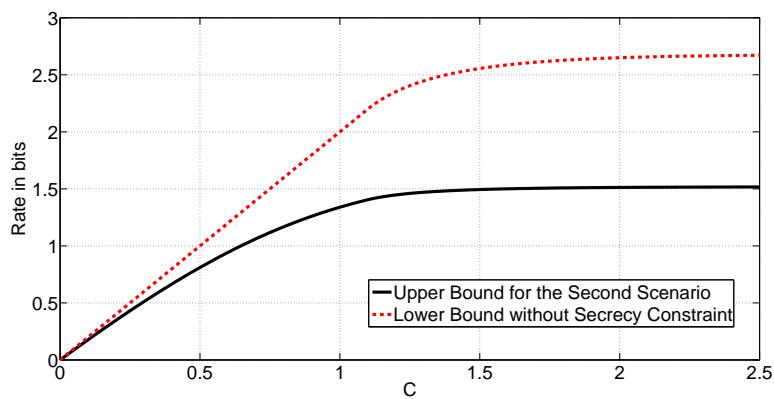}}\\
 \caption{Comparison with the lower and upper bounds without secrecy constraint when $P=10$ and $g=0.1$. } \label{fig:compare}
\end{figure}

Now, a natural question is how the presence of an eavesdropper affects the capacity. We partially answer this question by comparing our results with the lower and upper bounds in \cite{KangLiu:11} that are derived without secrecy constraint. Note that when there is no secrecy constraint, the availability of randomness at the encoders does not affect the capacity. Hence, the capacity without secrecy constraint is higher than or equal to the secrecy capacity with secrecy constraint both for the first and the second scenarios. We compare the bounds in Fig. \ref{fig:compare} for sufficiently large $R'$ and symmetric channel parameters. First, as illustrated in Fig. \ref{fig:compare}-(a), the upper bound without secrecy constraint and the lower bound for the first scenario coincide up to $C\leq \frac{1}{2}(f_4(0)-f_5(0))$. This indicates that, when there is a sufficient amount of common randomness between the source and the relays, there is no decrease in capacity due to an eavesdropper for some range of $C$. On the other hand, for the same channel parameters, Fig. \ref{fig:compare}-(b) shows that the lower bound without secrecy constraint is strictly higher than the upper bound for the second scenario for all range of $C>0$. This indicates that, when there is no  randomness at the relays, the secrecy capacity for the second scenario can be strictly smaller than the capacity without secrecy constraint for all range of $C$. 

\subsubsection*{Proof of Theorem \ref{thm:asymptotic}}
The bound in Theorem \ref{thm:common_ub} is further upper-bounded as follows: 
\begin{align*}
\min(\max(S_1, S_2), \max(S_3, S_4)) &\leq \max(S_3, S_4)\\
& \overset{(a)}{\leq} \max_{0\leq \rho \leq 1} \min(f_1(0), f_2(0), f_3(0), f_4(\rho)-f_5(\rho)),
\end{align*}
where $(a)$ is because $f_1(\rho)$ and $f_2(\rho)$ are decreasing functions of $\rho \in [0,1]$. Furthermore, for any $\rho\in [0,1]$, we have 
\begin{align*}
\lim_{P_1\rightarrow \infty \mbox{ or } P_2 \rightarrow \infty}f_4(\rho)-f_5(\rho) &=\lim_{P_1\rightarrow \infty \mbox{ or } P_2 \rightarrow \infty}\frac{1}{2}\log \frac{1+P_1+P_2+2\rho\sqrt{P_1P_2}}{1+g(P_1+P_2+2\rho\sqrt{P_1P_2})} \\
&=\frac{1}{2}\log \frac{1}{g}\\
&=\lim_{P_1\rightarrow \infty \mbox{ or } P_2 \rightarrow \infty}f_4(0)-f_5(0). 
\end{align*}
Hence, the secrecy capacity for the first scenario when $P_1\rightarrow \infty$ or $P_2\rightarrow \infty$ is asymtotically upper-bounded by 
\begin{align*}
\lim_{P_1\rightarrow \infty \mbox{ or } P_2 \rightarrow \infty}  \min(f_1(0), f_2(0), f_3(0), f_4(0)-f_5(0)), 
\end{align*}
which is clearly achievable by the PDF scheme. 
\endproof

\subsubsection*{Proof of Theorem \ref{thm:cap_condition}}
Let us first show that $\max (T_1,T_2)=f_3(\rho')-f_5(\rho')$. For symmetric channel parameters, $T_1$ and $T_2$ can be rewritten as follows: 
\begin{align*}
T_1&=\max_{-1\leq \rho < 0} \min(f_3(0), f_4(\rho))-f_5(\rho) \\
T_2&=\max_{0\leq \rho \leq \rho^*} \min(f_3(\rho), f_4(\rho))-f_5(\rho). 
\end{align*}
Let us define functions $f_{35}(\rho)$ and $f_{45}(\rho)$ of $\rho\in [-1, \rho^*]$ as follows: 
\begin{align*}
f_{35}(\rho)=\begin{cases}
f_3(0)-f_5(\rho) \mbox { if } -1\leq \rho <0 \\
f_3(\rho)-f_5(\rho) \mbox { otherwise}
\end{cases}, ~
f_{45}(\rho)=f_4(\rho)-f_5(\rho). 
\end{align*}
Note that we can rewrite the condition in (\ref{eqn:cap_condition}) as $f_{35}(0)\geq f_{45}(0)$ and $f_{35}(\rho^*)\leq f_{45}(\rho^*)$. Since  $f_{35}(\rho)$ and $f_{45}(\rho)$ are monotonically decreasing function and monotonically increasing function of $\rho\in [-1, \rho^*]$, respectively, the condition  in (\ref{eqn:cap_condition}) implies that  there exists $\rho'\in [0, \rho^*]$ such that $f_{35}(\rho')=f_{45}(\rho')$. Hence, we have $\max (T_1,T_2)=\max_{-1\leq \rho\leq \rho^*} \min(f_{35}(\rho), f_{45}(\rho))=f_{35}(\rho')=f_3(\rho')-f_5(\rho')$. 

Now, let us show $\max(f_{35}(\rho'),T_3)=f_{35}(\rho')$. Since both $f_1(\rho)-f_5(\rho)$ and $f_3(0)-f_5(\rho)$ for $\rho\in [\rho^*, 1]$ have the maximum at $\rho=\rho^*$,  we have 
\begin{align*}
\max(f_{35}(\rho'), T_3)\leq \max(f_{35}(\rho'), \min(f_1(\rho^*)-f_5(\rho^*), f_3(0)-f_5(\rho^*)))=f_{35}(\rho') 
\end{align*}
if $f_1(\rho^*)-f_5(\rho^*)\leq f_3(\rho')-f_5(\rho')$ or $f_3(0)-f_5(\rho^*)\leq f_3(\rho')-f_5(\rho')$. Hence, under the conditions in Theorem \ref{thm:cap_condition}, the upper bound in Theorem \ref{thm:no_ub} becomes $f_3(\rho')-f_5(\rho')$. 

Now, it remains to show $f_3(\rho')-f_5(\rho')$ is achievable. We have 
\begin{align*}
R_{\mathrm{PDF-PDF-M}}^{(2)}(\rho')&=(f_3(\rho')-f_5(\rho'))\cdot \mathbbm{1}_{C>f_6(\rho')}\\
&\overset{(a)}{=}f_3(\rho')-f_5(\rho')
\end{align*}
where $(a)$ is because $f_3(\rho')=f_4(\rho')$ and $f_3(\rho')-f_5(\rho')=f_4(\rho')-f_5(\rho')>0$ imply $C>f_6(\rho')$. This completes the proof.
\endproof

\section{Derivation of Upper and Lower Bounds on the Secrecy Capacity} \label{sec:proof}
In this section, we prove the upper and lower bounds on the secrecy capacity presented in Section \ref{sec:main}.
\subsection{Proof of Theorem \ref{thm:common_ub}} 
We note that the upper bound $\max(S_1,S_2)$, which the same as the upper bound in \cite{KangLiu:11} on the capacity  without secrecy constraint, is easily obtained by noting that the secrecy capacity is upper-bounded by the capacity without secrecy constraint and that common randomness at the encoders does not affect the capacity when there is no secrecy constraint. Nevertheless, we provide a direct proof for the upper bound $\max(S_1, S_2)$ as well as the upper bound $\max(S_3, S_4)$ since it can be useful for bounding in other related problems.  

The proof generalizes those in \cite{KangLiu:11} and \cite{TekinYener:08} taking into account the secrecy constraint and the available randomness at the encoders. For $k\in[1:2]$ and $i\in [1:n]$, let $P_{k,i}=\E(X_{k,i}^2)$ and let $\lambda_i=\frac{E(X_{1,i}X_{2,i})}{\sqrt{P_{1,i}P_{2,i}}}$. Let $\lambda_a\in [0,1]$ and $\lambda_b\in [0,1]$ be such that $\lambda_a^2P_1=\frac{1}{n}\sum_{i=1}^n \lambda_i^2P_{1,i}$ and $\lambda_b^2P_2=\frac{1}{n}\sum_{i=1}^n \lambda_i^2P_{2,i}$. We use $\epsilon_n$ to denote a function of $n$ such that $\epsilon_n$ tends to zero as $n$ tends to infinity.

By applying similar bounding techniques as in  \cite{KangLiu:11}, we have 
\begin{align}
nR&=H(W)\cr
&\overset{(a)}{\leq} I(W;J_1,Y^n,M)+n\epsilon_n\cr
&\overset{(b)}{=}I(W;J_1,Y^n|M)+n\epsilon_n\cr
&\leq H(J_1)+I(W;Y^n|J_1,M)+n\epsilon_n \cr
&\overset{(c)}{\leq} H(J_1)+I(W;Y^n|J_1,M,X_1^n)+n\epsilon_n \cr
&\leq H(J_1)+I(W,X_2^n;Y^n|J_1,M,X_1^n)+n\epsilon_n \cr
&\leq nC_1+I(X_2^n;Y^n|X_1^n)+n\epsilon_n\cr
&\leq nC_1+\sum_{i=1}^nI(X_{2,i};Y_{i}|X_{1,i})+n\epsilon_n\cr
&\overset{(d)}{\leq} nC_1+\sum_{i=1}^n\log (1+(1-\lambda_i^2)P_{2,i})+n\epsilon_n \cr
&\overset{(e)}{\leq}nC_1+n\log (\frac{1}{n}\sum_{i=1}^n(1+(1-\lambda_i^2)P_{2,i}))+n\epsilon_n \cr
&\overset{(f)}{\leq}nC_1+n\log (1+(1-\lambda_b^2)P_{2})+n\epsilon_n \label{eqn:f1} 
\end{align}
for sufficiently large $n$, where $(a)$ is from the Fano's inequality, $(b)$ is because $W$ and $M$ are independent, $(c)$ is because $X_1^n$ is a function of $J_1$ and $M$, $(d)$ is because the Gaussian distribution maximizes the differential entropy given the power constaint, $(e)$ is due to the concavity of the logarithm function, and $(f)$ is from the definition of $\lambda_b$. Similarly, we can obtain 
\begin{align}
nR&\leq   nC_2+n\log (1+(1-\lambda_a^2)P_{1})+n\epsilon_n \label{eqn:f2}
\end{align}
for sufficiently large $n$. 

We also have for sufficiently large $n$,
\begin{align}
nR&=H(W)\cr
&\overset{(a)}{\leq} I(W;Y^n,M)+n\epsilon_n\cr
&\overset{(b)}{=}I(W;Y^n|M)+n\epsilon_n\cr
&\overset{(c)}{=}I(X_1^n,X_2^n;Y^n|M)+n\epsilon_n \label{eqn:fourth}\\
&\leq H(X_1^n,X_2^n|M)+n\epsilon_n\cr
&\leq H(X_1^n|M)+H(X_2^n|M)-I(X_1^n;X_2^n|M)+n\epsilon_n\cr
&\leq nC_1+nC_2-I(X_1^n;X_2^n|M)+n\epsilon_n,  \label{eqn:fifth} 
\end{align}
where $(a)$ is from the Fano's inequality, $(b)$ is because $W$ and $M$ are independent, and $(c)$ is because $X_1^n$ and $X_2^n$ are functions of $M$ and $W$ and the Markov relationship $W-(M,X_1^n, X_2^n)-Y^n$ holds. 

Furthermore, for any random variable $U_i$ generated through a conditional pmf $p(u_i|x_{1,i}, x_{2,i}, y_i)$, we have  
\begin{align}
&I(X_1^n;X_2^n|M)\cr
&=\!I(X_1^n,X_2^n;U^n|M)-I(X_1^n;U^n|X_2^n,M)-I(X_2^n;U^n|X_1^n,M)+I(X_1^n;X_2^n|U^n,M)\cr
&\geq \!I(X_1^n,\!X_2^n;U^n|M)\!-\!I(X_1^n;U^n|X_2^n)\!-\!I(X_2^n;U^n|X_1^n). \label{eqn:intro_U}
\end{align}
By applying the above lower bound to (\ref{eqn:fifth}), we obtain 
\begin{align}
nR&\leq nC_1+nC_2-I(X_1^n,X_2^n;U^n|M)+I(X_1^n;U^n|X_2^n)+I(X_2^n;U^n|X_1^n)+n\epsilon_n. \label{eqn:f3}
\end{align}

For sufficiently large $n$, we have 
\begin{align}
nR&=H(W)\cr
&\overset{(a)}{\leq} H(W|Z^n)+n\epsilon_n \cr
&\overset{(b)}{\leq}  H(W|Z^n)-H(W|Y^n,Z^n)+2n\epsilon_n \cr
&= I(W;Y^n|Z^n)+2n\epsilon_n \cr
&\leq I(X_1^n,X_2^n;Y^n|Z^n)+2n\epsilon_n \cr
&\overset{(c)}{\leq} I(X_1^n,X_2^n;Y^n)-I(X_1^n,X_2^n;Z^n)+2n\epsilon_n \label{eqn:f4_pre}\\
&=h(Y^n)-h(Z^n)+2n\epsilon_n \cr
&\overset{(d)}{\leq} h(Y^n)-\frac{n}{2}\log (g2^{\frac{2}{n}h(Y^n)}+2\pi e (1-g))+2n\epsilon_n, \label{eqn:f4}
\end{align}
where $(a)$ is from the secrecy constraint, $(b)$ is due to the Fano's inequality, $(c)$ is due to the degradedness of the channel, and $(d)$ is from the entropy power inequality. We note that (\ref{eqn:f4}) is a nondecreasing function of $h(Y^n)$. $h(Y^n)$ is further upper-bounded as follows:
\begin{align*}
h(Y^n)&\leq \sum_{i=1}^n h(Y_i)\cr
&\leq \sum_{i=1}^n\frac{1}{2}\log (2\pi e)(1+P_{1,i}+P_{2,i}+2\lambda_i\sqrt{P_{1,i}P_{2,i}})\cr
&\leq \frac{n}{2}\log (2\pi e)(\frac{1}{n}\sum_{i=1}^n(1+P_{1,i}+P_{2,i}+2\lambda_i\sqrt{P_{1,i}P_{2,i}}))\cr
&\leq \frac{n}{2}\log (2\pi e)(1+P_{1}+P_{2}+\frac{2}{n}\sum_{i=1}^n\sqrt{\lambda_i^2P_{1,i}P_{2,i}}).
\end{align*}
From the Cauchy-Schwarz inequality, we have 
\begin{align}
\frac{1}{n}\sum_{i=1}^n\sqrt{\lambda_i^2P_{1,i}P_{2,i}}&\leq \sqrt{(\frac{1}{n}\sum_{i=1}^n \lambda_i^2 P_{1,i})(\frac{1}{n}\sum_{i=1}^nP_{2,i})}\cr
&\leq \sqrt{\lambda_a^2P_1P_2}.\nonumber
\end{align}
Similarly, we have 
$\frac{1}{n}\sum_{i=1}^n\sqrt{\lambda_i^2P_{1,i}P_{2,i}}\leq \sqrt{\lambda_b^2P_1P_2}$. 
Hence, we obtain
\begin{align}
h(Y^n)&\leq \frac{n}{2}\log (2\pi e)(1+P_{1}+P_{2}+2\min(\lambda_a, \lambda_b)\sqrt{P_1P_2}).  \label{eqn:upper_min}
\end{align}

Now we are ready to prove Theorem \ref{thm:common_ub}. Define $\mu\in [0,1]$ and $\nu\in [0,1]$ as follows. First,  $\mu$ is determined from $h(Y^n|M)$. $\mu=0$ if 
\begin{align}
\frac{1}{n}h(Y^n|M)\leq \frac{1}{2}\log (2\pi e)(1+P_1+P_2). \label{eqn:mu_zero}
\end{align}
Otherwise, $\mu$ is such that 
\begin{align}
\frac{1}{n}h(Y^n|M)=\frac{1}{2}\log (2\pi e)(1+P_1+P_2+2\mu\sqrt{P_1P_2}). \label{eqn:mu_nonzero}
\end{align}
Next, $\nu$ is determined from $h(Y^n)$. $\nu=0$ if
\begin{align}
\frac{1}{n}h(Y^n)\leq \frac{1}{2}\log (2\pi e)(1+P_1+P_2). \label{eqn:nu_zero}
\end{align}
 Otherwise, $\nu$ is such that 
\begin{align}
\frac{1}{n}h(Y^n)=\frac{1}{2}\log (2\pi e)(1+P_1+P_2+2\nu\sqrt{P_1P_2}).\label{eqn:nu_nonzero}
\end{align}

Let us first show that 
\begin{align}
R\leq \max(S_1,S_2)+\epsilon_n. \label{eqn:thm1_first}
\end{align}
If $\mu=0$, from (\ref{eqn:f1}), (\ref{eqn:f2}), (\ref{eqn:fifth}), (\ref{eqn:fourth}), and  (\ref{eqn:mu_zero}), we have $R\leq \min(f_1(0),f_2(0), f_3(0),f_4(0))+\epsilon_n$. Consider $\mu>0$. From $h(Y^n|M)\leq h(Y^n)$,  (\ref{eqn:mu_nonzero}), and (\ref{eqn:upper_min}), we have $\mu\leq \min(\lambda_a, \lambda_b).$ Then, from (\ref{eqn:f1}), (\ref{eqn:f2}), (\ref{eqn:fifth}),  (\ref{eqn:fourth}), and (\ref{eqn:mu_nonzero}), we obtain $R\leq \min(f_1(\mu),f_2(\mu), f_3(0),f_4(\mu))+\epsilon_n$. If $\mu$ further satisfies $0<\mu\leq \rho^*$, we let $U_i=Y_i+V_i$, where $V_i$ is an i.i.d. Gaussian random variable with zero mean and variance of $\gamma=\sqrt{P_1P_2}(\frac{1}{\mu}-\mu)-1$.\footnote{For $0<\mu\leq \rho^*$, $\gamma$ is nonnegative.} Then, the mutual information terms in (\ref{eqn:f3}) are bounded as follows: 
\begin{align}
&I(X_1^n,X_2^n;U^n|M)\cr
&\geq h(U^n|M)-\frac{n}{2}\log (2\pi e)(1+\gamma)\cr
&\overset{(a)}{\geq} \frac{n}{2} \log (2^{\frac{2}{n}h(Y^n|M)}+2\pi e\gamma)-\frac{n}{2}\log (2\pi e)(1+\gamma)\cr
&= \frac{n}{2} \log \frac{1+\gamma+P_1+P_2+2\mu\sqrt{P_1P_2}}{1+\gamma}\label{eqn:gamma1}\\
&I(X_1^n;U^n|X_2^n)\leq \frac{n}{2}\log\frac{1+\gamma+(1-\gamma^2)P_1}{1+\gamma} \label{eqn:gamma2}\\
&I(X_2^n;U^n|X_1^n)\leq \frac{n}{2}\log\frac{1+\gamma+(1-\gamma^2)P_2}{1+\gamma},\label{eqn:gamma3}
\end{align}
where $(a)$ is from the conditional entropy power inequality. Substituting the above bounds to (\ref{eqn:f3}), we obtain $R\leq f_3(\mu)+\epsilon_n$. Hence, we have $R\leq \min(f_1(\mu),f_2(\mu), f_3(\mu),f_4(\mu))+\epsilon_n$ for $0<\mu\leq \rho^*$. This concludes the proof of (\ref{eqn:thm1_first}).

Now, let us show  
\begin{align}
R\leq \max(S_3,S_4)+2\epsilon_n. \label{eqn:thm1_second}
\end{align}
If $\nu=0$, from (\ref{eqn:f1}), (\ref{eqn:f2}), (\ref{eqn:fifth}), (\ref{eqn:f4}), and (\ref{eqn:nu_zero}), we have $R\leq \min(f_1(0),f_2(0), f_3(0),f_4(0)-f_5(0))+2\epsilon_n$. Consider $\nu>0$. From  (\ref{eqn:nu_nonzero}) and (\ref{eqn:upper_min}), we have $\nu\leq \min(\lambda_a, \lambda_b).$ Then, from  (\ref{eqn:f1}), (\ref{eqn:f2}), (\ref{eqn:fifth}), (\ref{eqn:f4}), and (\ref{eqn:nu_nonzero}), we obtain $R\leq \min(f_1(\mu),f_2(\mu), f_3(0),f_4(\nu)-f_5(\nu))+2\epsilon_n$. If $\nu$ further satisfies $0<\nu\leq \rho^*$, we consider the following bound by adding the inequalities (\ref{eqn:fourth}) and (\ref{eqn:f3}): 
\begin{align}
2nR&\leq nC_1+nC_2+I(X_1^n,X_2^n;Y^n|M)-I(X_1^n,X_2^n;U^n|M)\cr
&~~~~~~~~~+I(X_1^n;U^n|X_2^n)+I(X_2^n;U^n|X_1^n) +2n\epsilon_n\cr
&\leq nC_1+nC_2+I(X_1^n,X_2^n;Y^n|U^n,M)\cr
&~~~~~~~~~+I(X_1^n;U^n|X_2^n)+I(X_2^n;U^n|X_1^n)+2n\epsilon_n\cr
&\leq nC_1+nC_2+I(X_1^n,X_2^n;Y^n|U^n)\cr
&~~~~~~~~~+I(X_1^n;U^n|X_2^n)+I(X_2^n;U^n|X_1^n)+2n\epsilon_n\cr
&\overset{(a)}{\leq} nC_1+nC_2+I(X_1^n,X_2^n;Y^n)-I(X_1^n,X_2^n;U^n)\cr
&~~~~~~~~~+I(X_1^n;U^n|X_2^n)+I(X_2^n;U^n|X_1^n)+2n\epsilon_n, \label{eqn:f3f4}
\end{align}
where $(a)$ holds when $(X_1^n, X_2^n)-Y^n-U^n$. We let $U_i=Y_i+V_i'$, where $V_i'$ is an i.i.d. Gaussian random variable with zero mean and variance of $\gamma'=\sqrt{P_1P_2}(\frac{1}{\nu}-\nu)-1$. 
Then, by substituting (\ref{eqn:nu_nonzero}) and similar bounds as in (\ref{eqn:gamma1})-(\ref{eqn:gamma3}) to  (\ref{eqn:f3f4}), we obtain $R\leq \frac{f_3(\nu)+f_4(\nu)}{2}+\epsilon_n$. Hence, we have $R\leq \min(f_1(\nu), f_2(\nu), f_3(0), \frac{f_3(\nu)+f_4(\nu)}{2}$, $f_4(\nu)-f_5(\nu))+2\epsilon_n$ for $0<\nu\leq \rho^*$. This concludes the proof of (\ref{eqn:thm1_second}).

\subsection{Proof of Theorem \ref{thm:common_lb}}
Let us first assume that the channel from the relays to the legitimate destination and the eavesdropper is a discrete memoryless channel with a conditional pmf $p(y,z|x_1,x_2)$. Fix $p(x_1,x_2)$ and let 
\begin{align}
R'=I(X_1,X_2;Z)-\delta(\epsilon). \label{eqn:ach_fictitious}
\end{align}
Fix $\epsilon>0$. We use $\delta(\epsilon)$ to denote a function of $\epsilon$ such that $\delta(\epsilon)$ tends to zero as $\epsilon$ tends to zero. 

In the DF scheme, the source sends the message to both relays, which requires $R<\min (C_1, C_2)$. Once the relays share both the message and the fictitious message, we can treat the channel from the relays to the legitimate destination and the eavesdropper as a classical wiretap channel \cite{Wyner:75}, \cite{CsiszarKorner:78}  with randomness of rate $R'$ in (\ref{eqn:ach_fictitious}) and hence the secrecy rate of $R<I(X_1, X_2;Y)-I(X_1,X_2;Z)$ is achievable. By combining two inequalities for $R$, we conclude the following secrecy rate is achievable: 
\begin{align}
\min (C_1, C_2, I(X_1,X_2;Y)-I(X_1,X_2;Z)). \label{eqn:DFD_1}
\end{align}

The PDF-M scheme is described in the following.  
\begin{itemize}
\item Codebook generation: We represent the message $w\in [1:2^{nR}]$ as the partial message pair $(w_1,w_2)\in [1:2^{nR_1}]\times [1:2^{nR_2}]$ for some $R_1\geq 0$ and $R_2\geq 0$ such that 
\begin{align}
R_1+R_2=R, \label{eqn:ach_eq1}
\end{align}
i.e., $W_k$ for $k\in [1:2]$ is uniformly distributed over $[1:2^{nR_k}]$ and $W_1$ and $W_2$ are independent. Consider $\tilde{R}_k\geq 0$ for $k\in [1:2]$. For each $k\in[1:2]$ and $(w_k,m,l_k)\in [1:2^{nR_k}]\times [1:2^{nR'}]\times [1:2^{n\tilde{R}_k}]$, generate $x_k^n(w_k,m,l_k)$ independently according to $\prod_{i=1}^np(x_{k,i})$. 
\item Encoding at the source: For message $(w_1,w_2)$ and fictitious message $m$, the source finds an  $(l_1,l_2)$ such that 
\begin{align*}
(x_1^n(w_1,m,l_1), x_2^n(w_2,m,l_2))\in \mathcal{T}_{\epsilon}^{(n)}. 
\end{align*} 
For $k\in [1:2]$, the source sends $(w_k,l_k)$ to relay $k$.  

\item Encoding at relay $k\in [1:2]$: Note that fictitious message $m$ is given at relay $k$. After receiving $(w_k,l_k)$ from the source, relay $k$ sends $x_k^n(w_k,m,l_k)$. 

\item Decoding at the legitimate destination: The legitimate destination finds $(\hat{w}_1,\hat{w}_2, \hat{m}, \hat{l}_1,\hat{l}_2)$ such that 
\begin{align*}
(x_1^n(\hat{w}_1,\hat{m},\hat{l}_1), x_2^n(\hat{w}_2,\hat{m},\hat{l}_2), y^n)\in \mathcal{T}_{\epsilon}^{(n)}. 
\end{align*} 
The legitimate destination declares that $(\hat{w}_1, \hat{w}_2)$ is the message. 

\item Error analysis: From the mutual covering lemma \cite{ElGamlvanderMeulen:81}, the encoding error at the source averaged over the codebooks tends to zero as $n$ tends to infinity if
\begin{align}
\tilde{R}_1+\tilde{R}_2>I(X_1;X_2)+\delta(\epsilon).\label{eqn:ach_covering}
\end{align} 
For $k\in [1:2]$, the transmission of $(w_k,l_k)$ from the source to relay $k$ requires 
\begin{align} 
R_k+\tilde{R}_k < C_k. \label{eqn:ach_sending}
\end{align}
From the standard error analysis, the decoding error at the legitimate destination averaged over the codebooks tends to zero as $n$ tends to infinity if 
\begin{align}
&R_1+\tilde{R}_1<I(X_1;Y|X_2)+I(X_1;X_2)-\delta(\epsilon)\label{eqn:ach_pac1}\\
&R_2+\tilde{R}_2<I(X_2;Y|X_1)+I(X_1;X_2)-\delta(\epsilon)\label{eqn:ach_pac2}\\
&R_1+R_2+R'+\tilde{R}_1+\tilde{R}_2<I(X_1,X_2;Y)+I(X_1;X_2)-\delta(\epsilon).\label{eqn:ach_pac3}
\end{align}

\item Secrecy analysis: We can show $\lim_{n\rightarrow \infty} \frac{1}{n}I(W;Z^n|\mathcal{C})\leq \delta(\epsilon)+\epsilon$ if (\ref{eqn:ach_fictitious}) and the following inequalities are satisfied.
\begin{align}
&\tilde{R}_1<I(X_1;Z|X_2)+I(X_1;X_2)-\delta(\epsilon)\label{eqn:ach_sec1}\\
&\tilde{R}_2<I(X_2;Z|X_1)+I(X_1;X_2)-\delta(\epsilon)\label{eqn:ach_sec2}\\
&R'+\tilde{R}_1+\tilde{R}_2<I(X_1,X_2;Z)+I(X_1;X_2)-\delta(\epsilon)\label{eqn:ach_sec3}
\end{align}
See Section \ref{sub:secrecy_analysis} for the detail. 
\end{itemize}

Therefore, there exists a sequence of codes such that $P_e^{(n)}$ tends to zero and $\frac{1}{n}I(W;Z^n)\leq \delta(\epsilon)+\epsilon$ as $n$ tends to infinity if (\ref{eqn:ach_fictitious}), (\ref{eqn:ach_eq1})-(\ref{eqn:ach_sec3}) are satisfied. By performing Fourier-Mozkin elimination to (\ref{eqn:ach_fictitious}), (\ref{eqn:ach_eq1})-(\ref{eqn:ach_sec3})  and by taking $\epsilon\rightarrow 0$, the PDF-M rate of 
\begin{align}
\min(C_1+I(X_2;Y|X_1), C_2+I(X_1;Y|X_2), C_1+C_2-I(X_1;X_2), I(X_1,X_2;Y)-I(X_1,X_2;Z)) \label{eqn:PDFMD_1}
\end{align}
is obtained. From the standard discretization procedure \cite{McEliece:77}, $R_{\mathrm{DF}}^{(1)}(\rho)$ and $R_{\mathrm{PDF-M}}^{(1)}(\rho)$ are obtained by evaluating (\ref{eqn:DFD_1}) and (\ref{eqn:PDFMD_1}) for the degraded Gaussian diamond-wiretap channel discussed in Section \ref{sec:model} and a jointly Gaussian distribution $p(x_1,x_2)$ such that $x_k$ for $k\in [1:2]$ has zero mean and variance of $P_k$ and the correlation coefficient between $X_1$ and $X_2$ is $\rho\in [-1,1]$.

\subsection{Proof of Theorem \ref{thm:no_ub}}
We note that the upper bound (\ref{eqn:f4_pre}) continues to hold when the fictitious message is given only at the source. Then, we have 
\begin{align}
nR&\leq I(X_1^n,X_2^n;Y^n)-I(X_1^n,X_2^n;Z^n)+n\epsilon_n \label{eqn:f4_sec}\\
&\leq I(X_1^n,X_2^n;J_1,Y^n)-I(X_1^n,X_2^n;Z^n)+n\epsilon_n\cr
&\leq H(J_1)+I(X_1^n,X_2^n;Y^n|J_1)-I(X_1^n,X_2^n;Z^n)+n\epsilon_n \cr
&\overset{(a)}{\leq} H(J_1)+I(X_2^n;Y^n|J_1,X_1^n)-I(X_1^n,X_2^n;Z^n)+n\epsilon_n \cr
&\leq nC_1+I(X_2^n;Y^n|X_1^n)-I(X_1^n,X_2^n;Z^n)+n\epsilon_n\cr
&\overset{(b)}{\leq} nC_1+n\log (1+(1-\lambda_b^2)P_{2})-I(X_1^n,X_2^n;Z^n)+n\epsilon_n \label{eqn:no_f1}
\end{align}
where  $\lambda_b$ is defined in the proof of Theorem \ref{thm:common_ub}, $(a)$ is because $X_1^n$ is a function of $J_1$, and $(b)$ is from some similar steps as in the derivation of  (\ref{eqn:f1}). Similarly, we can obtain 
\begin{align}
nR&\leq nC_2+n\log (1+(1-\lambda_a^2)P_{1})-I(X_1^n,X_2^n;Z^n)+n\epsilon_n, \label{eqn:no_f2}
\end{align}
where  $\lambda_a$ is defined in the proof of Theorem \ref{thm:common_ub}.

For any random variable $U_i$ generated through a conditional pmf $p(u_i|x_{1,i}, x_{2,i}, y_i)$, we have 
\begin{align}
nR&\leq I(X_1^n,X_2^n;Y^n)-I(X_1^n,X_2^n;Z^n)+n\epsilon_n \cr
&\leq H(X_1^n,X_2^n)-I(X_1^n,X_2^n;Z^n)+n\epsilon_n\cr
&\leq H(X_1^n)+H(X_2^n)-I(X_1^n;X_2^n)-I(X_1^n,X_2^n;Z^n)+n\epsilon_n\cr
&\overset{(a)}{\leq} nC_1+nC_2-I(X_1^n;X_2^n)-I(X_1^n,X_2^n;Z^n)+n\epsilon_n \label{eqn:four}\ \\
&\overset{(b)}{\leq}nC_1+nC_2-I(X_1^n,X_2^n;U^n)+I(X_1^n;U^n|X_2^n)\cr
&~~~~~~~~~~+I(X_2^n;U^n|X_1^n)-I(X_1^n,X_2^n;Z^n)+n\epsilon_n, \label{eqn:f4f5}
\end{align}
where $(a)$ is because $X_k^n$ is a function of $J_k$ for $k\in [1:2]$ and $(b)$ is from some similar steps as in the derivation of (\ref{eqn:intro_U}).

Note that we have the following lower and upper bounds on $\frac{1}{n}h(Y^n)$:
\begin{align*}
\frac{1}{n}h(Y^n)&\geq \frac{1}{n}h(Y^n|X_1^n, X_2^n)= \frac{1}{n}h(N_Y^n) = \frac{1}{2}\log (2\pi e) \\
\frac{1}{n}h(Y^n)&\leq \frac{1}{2}\log (2\pi e)(1+P_1+P_2+2\sqrt{P_1P_2}).
\end{align*}
Hence, there exists $\rho\in [-\bar{\rho}, 1]$ such that 
\begin{align}
\frac{1}{n}h(Y^n)=\frac{1}{2}\log (2\pi e)(1+P_1+P_2+2\rho\sqrt{P_1P_2}).\label{eqn:rho_def}
\end{align}
Then, we have the following lower bound on $I(X_1^n,X_2^n;Z^n)$:
\begin{align}
I(X_1^n,X_2^n;Z^n)
\geq \frac{n}{2}\log(1+g(P_1+P_2+2\rho\sqrt{P_1P_2})) \label{eqn:secrety_minus}
\end{align}
from the entropy power inequality. 

Now, we are ready to prove Theorem \ref{thm:no_ub}. First consider $\rho\in [-\bar{\rho},0)$. Then, from (\ref{eqn:f4_sec})-(\ref{eqn:four}), (\ref{eqn:rho_def}), and (\ref{eqn:secrety_minus}), we have $R\leq \min(f_1(0), f_2(0), f_3(0), f_4(\rho))-f_5(\rho)+\epsilon_n$. Next, consider $\rho\in [0,1]$. Then, due to similar reasons as in the proof of Theorem \ref{thm:common_ub}, we have $\rho\leq \min (\lambda_a, \lambda_b)$. Then, from (\ref{eqn:f4_sec})-(\ref{eqn:four}), (\ref{eqn:rho_def}), and (\ref{eqn:secrety_minus}), we have $R\leq \min(f_1(\rho), f_2(\rho), f_3(0), f_4(\rho))-f_5(\rho)+\epsilon_n$. Now, assume that $\rho$ further satisfies $\rho\in [0, \rho^*].$ We choose  $U_i=Y_i+\tilde{V}_i$, where $\tilde{V}_i$ is an i.i.d. Guassian random variable with zero mean and variance of  $\tilde{\gamma}=\sqrt{P_1P_2}(\frac{1}{\rho}-\rho)-1$. Then, by substituting (\ref{eqn:secrety_minus}) and similar bounds as (\ref{eqn:gamma1})-(\ref{eqn:gamma3}) to (\ref{eqn:f4f5}), we obtain $R\leq f_3(\rho)-f_5(\rho)+\epsilon_n$. Hence, we have $R\leq \min(f_1(\rho),f_2(\rho), f_3(\rho),f_4(\rho))-f_5(\rho)+\epsilon_n$ for $\rho\in [0, \rho^*]$. This concludes the proof of Theorem  \ref{thm:no_ub}.

\subsection{Proof of Theorem \ref{thm:no_lb}} 
As in the proof of Theorem \ref{thm:common_lb}, we first assume that the channel from the relays to the legitimate destination and the eavesdropper is a discrete memoryless channel with a conditional pmf $p(y,z|x_1,x_2)$.  Fix $p(x_1,x_2)$ and $\epsilon>0$.  Let
\begin{align}
R'=I(X_1,X_2;Z)-\delta(\epsilon). \label{eqn:ach_fictitious2}
\end{align}
For the DF scheme, by letting the source send both the message and the fictitious message to the relays, an achievable secrecy rate of  
\begin{align}
\min (C_1-R', C_2-R', I(X_1,X_2;Y)-I(X_1,X_2;Z)) \label{eqn:DFD_2}
\end{align}
is obtained from (\ref{eqn:DFD_1}) by replacing $C_1$ and $C_2$ by $C_1-R'$ and $C_2-R'$, respectively.

Similarly, for the PDF-DF-M scheme, by letting the source send the fictitious message as well as the partial message and the relay codeword index to relay $k$ for $k=1,2$ in the PDF-M scheme for the first scenario, an achievable secrecy rate of 
\begin{align}
\min(C_1+I(X_2;Y|X_1)-R',&~ C_2+I(X_1;Y|X_2)-R', \cr
&C_1+C_2-I(X_1;X_2)-2R', I(X_1,X_2;Y)-I(X_1,X_2;Z)) \label{eqn:PDFMD_2}
\end{align}
is obtained from (\ref{eqn:PDFMD_1}) by replacing $C_1$ and $C_2$ by $C_1-R'$ and $C_2-R'$, respectively.

The PDF-PDF-M scheme is described in the following.  
\begin{itemize}
\item Codebook generation: We represent the message $w\in [1:2^{nR}]$ and the fictitious message $m\in [1:2^{nR'}]$ as a partial message pair $(w_1,w_2)\in [1:2^{nR_1}]\times [1:2^{nR_2}]$ and  a partial fictitious  message pair $(m_1,m_2)\in [1:2^{nR_1'}]\times [1:2^{nR_2'}]$, respectively, for some nonnegative rates $R_1, R_2, R_1'$, and $R_2'$ such that 
\begin{align}
R_1+R_2=R, ~R_1'+R_2'=R'. \label{eqn:ach_eq2}
\end{align}
Consider $\tilde{R}_k\geq 0$ for $k\in [1:2]$. For each $k\in[1:2]$ and $(w_k,m_k,l_k)\in [1:2^{nR_k}]\times [1:2^{nR'_k}] \times [1:2^{n\tilde{R}_k}]$, generate $x_k^n(w_k,m_k,l_k)$ independently according to $\prod_{i=1}^np(x_{k,i})$. 
\item Encoding at the source: For message $(w_1,w_2)$ and fictitious message $(m_1,m_2)$, the source finds an  $(l_1,l_2)$ such that 
\begin{align*}
(x_1^n(w_1,m_1,l_1), x_2^n(w_2,m_2,l_2))\in \mathcal{T}_{\epsilon}^{(n)}. 
\end{align*} 
For $k\in [1:2]$, the source sends $(w_k,m_k,l_k)$ to relay $k$.  

\item Encoding at relay $k\in [1:2]$: After receiving $(w_k,m_k,l_k)$ from the source, relay $k$ sends $x_k^n(w_k,m_k,l_k)$. 

\item Decoding at the legitimate destination: The legitimate destination finds $(\hat{w}_1,\hat{w}_2,\hat{m}_1, \hat{m}_2,\hat{l}_1,\hat{l}_2)$ such that 
\begin{align*}
(x_1^n(\hat{w}_1,\hat{m}_1,\hat{l}_1), x_2^n(\hat{w}_2,\hat{m}_2,\hat{l}_2), y^n)\in \mathcal{T}_{\epsilon}^{(n)}. 
\end{align*} 
The legitimate destination declares $(\hat{w}_1, \hat{w}_2)$ is the message. 

\item Error analysis: From the mutual covering lemma, the encoding error at the source averaged over the codebooks tends to zero as $n$ tends to infinity if
\begin{align}
\tilde{R}_1+\tilde{R}_2>I(X_1;X_2)+\delta(\epsilon).\label{eqn:ach_covering}
\end{align} 
For $k\in [1:2]$, the transmission of $(w_k,m_k,l_k)$ from the source to relay $k$ requires 
\begin{align} 
R_k+R_k'+\tilde{R}_k < C_k. \label{eqn:ach_sending}
\end{align}
From the standard error analysis, the decoding error at the legitimate destination averaged over the codebooks tends to zero as $n$ tends to infinity if 
\begin{align}
&R_1+R_1'+\tilde{R}_1<I(X_1;Y|X_2)+I(X_1;X_2)-\delta(\epsilon)\label{eqn:ach_pac1_no}\\
&R_2+R_2'+\tilde{R}_2<I(X_2;Y|X_1)+I(X_1;X_2)-\delta(\epsilon)\label{eqn:ach_pac2_no}\\
&R_1+R_2+R_1'+R_2'+\tilde{R}_1+\tilde{R}_2<I(X_1,X_2;Y)+I(X_1;X_2)-\delta(\epsilon).\label{eqn:ach_pac3_no}
\end{align}

\item Secrecy analysis: We can show $\lim_{n\rightarrow \infty} \frac{1}{n}I(W;Z^n|\mathcal{C})\leq \delta(\epsilon)+\epsilon$ if (\ref{eqn:ach_fictitious2}) and the following inequalities are satisfied.
\begin{align}
&R_1'+\tilde{R}_1<I(X_1;Z|X_2)+I(X_1;X_2)-\delta(\epsilon)\label{eqn:ach_sec1_no}\\
&R_2'+\tilde{R}_2<I(X_2;Z|X_1)+I(X_1;X_2)-\delta(\epsilon)\label{eqn:ach_sec2_no}\\
&R_1'+R_2'+\tilde{R}_1+\tilde{R}_2<I(X_1,X_2;Z)+I(X_1;X_2)-\delta(\epsilon)\label{eqn:ach_sec3_no}
\end{align}
See Section \ref{sub:secrecy_analysis} for the detail. 
\end{itemize}

Therefore, there exists a sequence of codes such that $P_e^{(n)}$ tends to zero and $\frac{1}{n}I(W;Z^n)\leq \delta(\epsilon)+\epsilon$ as $n$ tends to infinity if (\ref{eqn:ach_fictitious2}), (\ref{eqn:ach_eq2})-(\ref{eqn:ach_sec3_no}) are satisfied. By performing Fourier-Mozkin elimination to (\ref{eqn:ach_fictitious2}), (\ref{eqn:ach_eq2})-(\ref{eqn:ach_sec3_no}) and by taking $\epsilon\rightarrow 0$, a secrecy rate of 
\begin{align}
\min(C_1+I(X_2;Y|X_1), C_2+I(X_1;Y|X_2), C_1+C_2-I(X_1;X_2), I(X_1,X_2;Y))-I(X_1,X_2;Z) \label{eqn:PDFMD2}
\end{align}
subject to the constraints 
\begin{align}
C_1>I(X_1;Z), C_2>I(X_2;Z) \label{eqn:PDFMD2_constr}
\end{align}
is obtained. From the standard discretization procedure, $R_{\mathrm{DF}}^{(2)}(\rho)$, $R_{\mathrm{PDF-DF-M}}^{(2)}(\rho)$, and $R_{\mathrm{PDF-PDF-M}}^{(2)}(\rho)$ are obtained by evaluating (\ref{eqn:DFD_2}), (\ref{eqn:PDFMD_2}), (\ref{eqn:PDFMD2}), and (\ref{eqn:PDFMD2_constr}) for the degraded Gaussian diamond-wiretap channel discussed in Section \ref{sec:model} and a jointly Gaussian distribution $p(x_1,x_2)$ such that $x_k$ for $k\in [1:2]$ has zero mean and variance of $P_k$ and the correlation coefficient between $X_1$ and $X_2$ is $\rho\in [-1,1]$.

\subsection{Secrecy analysis} \label{sub:secrecy_analysis}
Let $\mathcal{C}$ denote the random codebook. For message $W$, fictitious message $M$, and chosen relay codeword indices $L=(L_1,L_2)$, we have
\begin{align*}
H(W|Z^n,\mathcal{C})&=H(W,M,L|Z^n,\mathcal{C})-H(M,L|W,Z^n,\mathcal{C}) \\
&\overset{(a)}{\geq} H(W,M, L|Z^n,\mathcal{C})-n\epsilon \\
&=H(W,M,L|\mathcal{C})-I(W,M,L;Z^n|\mathcal{C})-n\epsilon \\
&\geq H(W)+nR'-I(W, M, L, X_1^n,X_2^n,\mathcal{C};Z^n)-n\epsilon \\
&=H(W)+nR'-I(X_1^n,X_2^n;Z^n)-n\epsilon \\
&\geq H(W)+nR'-nI(X_1,X_2;Z)-n\epsilon\\
&= H(W)-n\delta(\epsilon)-n\epsilon
\end{align*}
for sufficiently lage $n$, where $(a)$ is because the eavesdropper who already knows  $W$ and $Z^n$ can decode $M$ and $L$ with high probability when (\ref{eqn:ach_sec1})-(\ref{eqn:ach_sec3}) are satisfied for the first scenario and when  (\ref{eqn:ach_sec1_no})-(\ref{eqn:ach_sec3_no}) are satisfied for the second scenario. Hence, we have $\lim_{n\rightarrow \infty} \frac{1}{n}I(W;Z^n|\mathcal{C})\leq \delta(\epsilon)+\epsilon$. 
\section{Conclusion} \label{sec:conclusion}
In this paper,  we derived nontrivial upper and lower bounds on the secrecy capacity of the degraded Gaussian diamond-wiretap channel under two scenarios regarding the availability of randomness. 

Our upper bound was obtained by taking into account the correlation between the two relay signals and the availability of randomness at each encoder, which generalizes both the upper bound on the capacity of the diamond channel without secrecy constraint \cite{KangLiu:11} and the upper bound on the sum secrecy capacity of the MAC wiretap channel \cite{TekinYener:08}.  For the lower bound, we proposed DF scheme and partial DF scheme incorporated with multicoding that is called PDF-M scheme for the first scenario and PDF-DF-M and PDF-PDF-M schemes for the second scenario depending on whether the relay decodes the whole or partial fictitious message. In the first scenario, PDF-M scheme with strictly positive correlation coefficient was shown to outperform DF and PDF (without multicoding) schemes  for some channel parameters. We also showed that the PDF scheme is asymptotically optimal for the first scenario when at least one of relay power constraint tends to infinity. For the second scenario, we
  presented a condition for channel parameters where the PDF-PDF-M scheme is optimal. Furthermore, because the fictitious message has to be sent through the BC for the second scenario, it was shown to be befinicial to consider negative correlation in all DF, PDF-DF-M, PDF-PDF-M schemes when the BC cut becomes the bottleneck. Furthermore, we investigated the effect of the presence of an eavesdropper on the capacity. If there is a sufficient amount of common randomness between the source and the relays, it was shown that there is no decrease in capacity due to an eavesdropper for some range of $C$. 
  
 As a final remark, it seems to be straightforward to combine our DF scheme and partial DF scheme incorporated with multicoding by using superposition coding, but the resultant rate expression would be rather complicated with less useful insights.

%

\begin{thebibliography}{10}
\providecommand{\url}[1]{#1}
\csname url@samestyle\endcsname
\providecommand{\newblock}{\relax}
\providecommand{\bibinfo}[2]{#2}
\providecommand{\BIBentrySTDinterwordspacing}{\spaceskip=0pt\relax}
\providecommand{\BIBentryALTinterwordstretchfactor}{4}
\providecommand{\BIBentryALTinterwordspacing}{\spaceskip=\fontdimen2\font plus
\BIBentryALTinterwordstretchfactor\fontdimen3\font minus
  \fontdimen4\font\relax}
\providecommand{\BIBforeignlanguage}[2]{{%
\expandafter\ifx\csname l@#1\endcsname\relax
\typeout{** WARNING: IEEEtran.bst: No hyphenation pattern has been}%
\typeout{** loaded for the language `#1'. Using the pattern for}%
\typeout{** the default language instead.}%
\else
\language=\csname l@#1\endcsname
\fi
#2}}
\providecommand{\BIBdecl}{\relax}
\BIBdecl

\bibitem{schein_thesis}
B.~E. Schein, ``Distributed coordination in network information theory,'' Ph.D.
  dissertation, Massachusetts Institute of Technology, 2001.

\bibitem{TraskovKramer:07}
D.~Traskov and G.~Kramer, ``Reliable communication in networks with
  multi-access interference,'' in \emph{Proc. {IEEE} Information Theory
  Workshop (ITW)}, 2007, pp. 343--348.

\bibitem{KangLiu:11}
W.~Kang and N.~Liu, ``The {G}aussian multiple access diamond channel,'' in
  \emph{Proc. {IEEE} Int. Symp. Inform. Theory (ISIT)}, Jul.-Aug. 2011, pp.
  1499--1503.

\bibitem{TekinYener:08}
E.~Tekin and A.~Yener, ``The {G}aussian multiple access wire-tap channel,''
  \emph{{IEEE} Trans. Inf. Theory}, vol.~54, pp. 5747--5755, Dec. 2008.

\bibitem{Kocher:99}
P.~Kocher, J.~Jaffe, and B.~Jun, ``Differential power analysis,'' in
  \emph{Proc. Advances in Cryptology (CRYPTO '99)}, 1999, pp. 388--397.

\bibitem{KobayashiYamamotoOgawa:11}
D.~Kobayashi, H.~Yamamoto, and T.~Ogawa, ``How to attain the ordinary channel
  capacity securely in wiretap channels,'' in \emph{Proc. {IEEE} Information
  Theory Workshop on Theory and Practice in Information-Theoretic Security},
  Oct. 2005, pp. 13--18.

\bibitem{ChouBlock:14}
R.~A. Chou and M.~R. Bloch, ``Uniform distributed source coding for the
  multiple access wiretap channel,'' in \emph{Proc. {IEEE} Conference on
  Communications and Network Security (CNS)}, Oct. 2014, pp. 127--132.

\bibitem{Wyner:75}
A.~D. Wyner, ``The wire-tap channel,'' \emph{Bell Syst. Tech. J.}, vol.~54, pp.
  1355--1387, 1975.

\bibitem{CsiszarKorner:78}
I.~Csisz{\'{a}}r and J.~K{\"{o}}rner, ``Broadcast channels with confidential
  messages,'' \emph{{IEEE} Trans. Inf. Theory}, vol.~24, pp. 339--348, May
  1978.

\bibitem{ElGamlvanderMeulen:81}
A.~{El Gamal} and E.~C. {van der Meulen}, ``A proof of {M}arton's coding
  theorem for the discrete memoryless broadcast channel,'' \emph{{IEEE} Trans.
  Inf. Theory}, vol.~27, pp. 120--122, Jan. 1981.

\bibitem{McEliece:77}
R.~J. McEliece, \emph{The theory of information and coding}.\hskip 1em plus
  0.5em minus 0.4em\relax Addison-Wesley, Reading, 1977.

\end{thebibliography}

\end{document}